\documentclass[12pt,letterpaper]{article}%
\usepackage{amsmath,amssymb,amsfonts}
\usepackage{graphicx,epsfig,url}
\usepackage{color}
\usepackage{float}
\usepackage{caption2}
\usepackage{hyperref}
\usepackage[Symbolsmallscale]{upgreek}
\usepackage{amsmath}
\usepackage{amsfonts}
\usepackage{amssymb}
\usepackage{graphicx}%
\numberwithin{equation}{section} 
\setcaptionmargin{1cm}

\font\tenrsfs=rsfs10 at 12pt \font\sevenrsfs=rsfs7
\font\fiversfs=rsfs5
\newfam\rsfsfam
\textfont\rsfsfam=\tenrsfs
\scriptfont\rsfsfam=\sevenrsfs
\scriptscriptfont\rsfsfam=\fiversfs
\def\mathscr#1{{\fam\rsfsfam\relax#1}}
\begin{document}
\begin{titlepage}
\begin{flushright}
\end{flushright}
\vskip 1.0cm
\begin{center}
{\Large \bf Scale Invariance $+$ Unitarity $\Longrightarrow$
Conformal Invariance?} \vskip 1.0cm {\large Daniele Dorigoni$^{a,b}$
and
Slava Rychkov$^{c,a}$} \\[0.7cm]
{\it $^a$ Scuola Normale Superiore and INFN, Sezione di Pisa, Italy}\\
{\it $^b$ Dipartimento di Fisica, Universit\`a di Pisa, Italy}\\
{\it $^c$
Laboratoire de Physique Th\'{e}orique, Ecole Normale Superieure,\\
and Facult\'{e} de physique, Universit\'{e} Paris VI, France}
\end{center}
\begin{abstract}
We revisit the long-standing conjecture that in unitary field
theories, scale invariance implies conformality. We explain why the
Zamolodchikov-Polchinski proof in $D=2$ does not work in higher
dimensions. We speculate which new ideas might be helpful in a
future proof. We also search for possible counterexamples. We
consider a general multifield scalar-fermion theory with quartic and
Yukawa interactions. We show that there are no counterexamples among
fixed points of such models in $4-\varepsilon$ dimensions. We also
discuss fake counterexamples, which exist among theories without a
stress tensor.
\end{abstract}
\vskip 2cm \hspace{0.7cm} October 2009
\end{titlepage}

\section{Introduction}

Conformal invariance is a fundamental concept in statistical mechanics, high
energy physics, and string theory. It is usually considered to be a natural
consequence of scale invariance. However, the precise relation between these
two spacetime symmetries is more subtle.

In $D=2$ spacetime dimensions the situation is well understood, as we have a
theorem \cite{Zamolodchikov:1986gt},\cite{Polchinski:1987dy}:
\textit{any\footnote{There are two mild technical assumptions, discussed in
Section \ref{sec:2D} below.} scale invariant, unitary 2D QFT is conformally
invariant.} The assumption of unitarity is essential, as the example of the
free vector field without gauge invariance shows.

It $D\geq3$, it is not known if the above theorem is valid. A proof has never
been given, and there is no known reason why it should be generally true. At
the same time, there is no known counterexample. Given the importance of
conformal invariance, the situation is rather embarrassing. It is also in
contradiction to Gell-Mann's ``Everything which is not forbidden, is compulsory."

As our title shows, we would like to reopen the discussion of this interesting
problem\footnote{See also \cite{Grinstein:2008qk},\cite{Nakayama:2009ww}%
,\cite{Nakayama:2009qu} for recent mentionings of the issue.
Ref.\ \cite{Nakayama:2009fe} has shown that, at the level of string/M-theory
low energy effective actions, it is impossible to deform AdS/CFT by breaking
conformal invariance while preserving scale invariance. }. The paper is
organized as follows.

In Section \ref{sec:2pt} we revisit the proof in $D=2$. Very little dynamical
information is used in this well-known proof. It is enough to study the
2-point function of the symmetric stress tensor $T_{\mu\nu}$, which is fixed
by Lorentz and scale invariance up to a few numerical coefficients. The stress
tensor conservation provides a strong constraint on these coefficients, to the
extent that a particular linear combination which enters the 2-point function
of the trace $T_{\mu}^{\mu}$ is constrained to be zero. It is at this point
that the unitarity is invoked to conclude that $T_{\mu}^{\mu}=0$, and thus the
theory is conformally invariant.

We then discuss why this $D=2$ proof does not generalize to higher dimensions.
The conclusions of Section 2 is that if a proof in $D\geq3$ exists, it must be
based on ideas essentially different from $D=2$. We speculate that perhaps the
stress tensor Ward identities may turn out useful.

Having failed to find a proof, we proceed to look for counterexamples. For any
$D$, the scale current of the theory is related to $T_{\mu\nu}$ via
\cite{Wess:1960}:
\begin{equation}
S_{\mu}(x)=x^{\nu}T_{\mu\nu}(x)-K_{\mu}(x)\,,\label{scalecurrent}%
\end{equation}
where $K^{\mu}(x)$ is a local operator without explicit dependence on the
coordinates. In a scale invariant theory the current (\ref{scalecurrent}) is
conserved, which means:
\begin{equation}
T_{\mu}^{\mu}=\partial_{\mu}K^{\mu}\,.\label{eq:TK}%
\end{equation}
Now, if $K_{\mu}$ has the special form%
\begin{equation}
K^{\mu}=\partial_{\nu}L^{\nu\mu},\label{eq:KL}%
\end{equation}
where $L_{\mu\nu}$ is another local field, then a conserved conformal current
can be constructed. At the same time, one can construct an `improved'
symmetric stress tensor $T_{\mu\nu}^{\prime}$ which is traceless:%
\begin{equation}
T_{\mu}^{\prime\mu}=0\,.\label{eq:theta}%
\end{equation}
$T_{\mu\nu}^{\prime}$ is obtained from $T_{\mu\nu}$ by adding a local operator
which is a total divergence and identically conserved:%
\begin{equation}
T_{\mu\nu}^{\prime}=T_{\mu\nu}+\partial^{\rho}\partial^{\sigma}Y_{\mu\rho
\nu\sigma}\,.\label{eq:EMmodif}%
\end{equation}
Here $Y_{\mu\rho\nu\sigma}$ is antisymmetric in $\mu\rho$ and $\nu\sigma$ and
symmetric under $\mu\rho\leftrightarrow\nu\sigma$; such an operator can be
constructed in terms of $L_{\mu\nu}$ \cite{Polchinski:1987dy}. Thus $T_{\mu
\nu}^{\prime}$ is physically equivalent to $T_{\mu\nu}$, and the condition
(\ref{eq:theta})\ makes conformal symmetry manifest.

As stressed by Polchinski \cite{Polchinski:1987dy}, this analysis narrows
significantly the circle of possible counterexamples. Namely, a theory must
have nontrivial candidates for a dimension 3 vector operator $K_{\mu}$ which
is a) not conserved and b) is not of the form (\ref{eq:KL}). Several
better-known perturbative fixed points, such as the Belavin-Migdal-Banks-Zaks
fixed points for non-abelian gauge theories coupled to fermions in $D=4$
\cite{belavin}, or the Wilson-Fisher $\lambda\phi^{4}$ fixed point in
$D=4-\varepsilon$ \cite{Wilson} do not contain such a candidate, and thus are
automatically conformally invariant.

The simplest class of theories with $K_{\mu}$ candidates are the multi-field
generalizations of $\lambda\phi^{4}$:%
\begin{equation}
\mathcal{L}=\frac{1}{2}\left(  \partial s_{i}\right)  ^{2}+\frac{1}{4!}%
\lambda_{ijkl}s_{i}s_{j}s_{k}s_{l}\,. \label{eq:phi4}%
\end{equation}
The $K_{\mu}$ could be given by
\begin{equation}
K_{\mu}=Q_{[ij]}\,s_{i}\partial_{\mu}s_{j}\,, \label{eq:Kmu}%
\end{equation}
where $Q_{[ij]}$ is a real antisymmetric matrix. These theories can have
perturbative fixed points in $D=4-\varepsilon$. The usual fixed points,
obtained by setting the one-loop $\beta$-functions to zero, have $T_{\mu}%
^{\mu}=0$ and are conformally invariant. One could hope that more general
fixed points of the type (\ref{eq:TK}) exist. However, this turns out not to
be the case \cite{Polchinski:1987dy}. Namely, one can show that if
(\ref{eq:TK}) holds, then necessarily $K_{\mu}=0$, and we are back to the
usual case.

To our knowledge, this was the first and the last systematic attempt to look
for counterexamples, and it did not succeed. We don't really know why: the
only known proof is by inspection. Nor do we know what happens in higher
orders of perturbation theory. In our opinion, it is important to continue the
search. Clearly, (\ref{eq:phi4}) is not the most general unitary field theory
which can have fixed points in $D=4-\varepsilon$.

Thus, in Section \ref{sec:count} we consider a more general model which
contains an arbitrary number of scalars and Weyl fermions, with quartic and
Yukawa interactions. The full Lagrangian consists of (\ref{eq:phi4}) and of%
\begin{equation}
\mathcal{L}^{\prime}=i{\bar{\psi}}_{a}{\bar{\sigma}}_{\mu}\partial_{\mu}%
\psi_{a}+\frac{1}{2!}(y_{i|ab}s_{i}\psi_{a}\psi_{b}+y_{i|ab}^{\ast}s_{i}%
{\bar{\psi}}_{a}{\bar{\psi}}_{b})\,. \label{eq:psi}%
\end{equation}
We thus allow both for possible parity and CP breaking. It's not clear if
these discrete symmetries have anything to do with the relation between scale
and conformal invariance, but we would like to explore without prejudice.

We then look for one-loop fixed points with scale but without conformal
invariance. Once again, it turns out that there are none. The proof of this
fact is however more involved than in the case without fermions. We find a way
to represent it graphically: contractions of the $\beta$-function Feynman
graphs with graphs representing $K_{\mu}$ candidates all vanish by
(anti)symmetry. Unfortunately, this is still a proof by inspection: we don't
explain \textit{why} all contracted graphs had the needed symmetry properties.
This is a problem for the future.

In all of the above discussion we were assuming, for good reasons indeed, that
a stress tensor exists. If one drops this physical requirement, there are
theories which are unitary and scale invariant but not conformal. We mention a
couple of such \textit{fake} counterexamples in Section \ref{sec:count}.

We conclude in Section \ref{concl} with comments about the possible impact of
the eventual resolution of this problem on the TeV-scale phenomenology.

\section{Looking for a proof}

\label{sec:2pt}

Zamolodchikov had an idea to take the stress tensor 2-point function, impose
conservation, and see what comes out of it. What comes out is the 2D
$c$-theorem \cite{Zamolodchikov:1986gt}, and the fact that any 2D scale
invariant theory is conformally invariant \cite{Polchinski:1987dy}.
Unfortunately, the trick works only in $D=2$. In this section we would like to
discuss why this is so. If the theorem is valid in $D\geq3$, some extra ideas
should appear in the proof, and we outline one such preliminary idea.

\subsection{Theorem in $D=2$}

\label{sec:2D}

We will present here a slightly different version of the proof
\cite{Zamolodchikov:1986gt},\cite{Polchinski:1987dy} that in $D=2$ scale
invariance plus unitarity implies conformal invariance.


Consider the 2-point function of the symmetric stress
tensor\footnote{Eq.~(4.72) in Section 4.3.3 of \cite{Difran} is missing the
$t_{3}$ term.}%
\begin{align}
\left\langle T_{\mu\nu}(x)T_{\lambda\sigma}(0)\right\rangle  &  =[t_{1}%
\,(\eta_{\mu\lambda}\eta_{\nu\sigma}+\eta_{\nu\lambda}\eta_{\mu\sigma}%
)+t_{2}\,\eta_{\mu\nu}\eta_{\lambda\sigma}]/(x^{2})^{2}\nonumber\\
&  +t_{3}\,(\eta_{\mu\lambda}x_{\nu}x_{\sigma}+\eta_{\mu\sigma}x_{\nu
}x_{\lambda}+\eta_{\nu\lambda}x_{\mu}x_{\sigma}+\eta_{\nu\sigma}x_{\mu
}x_{\lambda})/(x^{2})^{3}\nonumber\\
&  +t_{4}\,(\eta_{\mu\nu}x_{\lambda}x_{\sigma}+\eta_{\lambda\sigma}x_{\mu
}x_{\nu})/(x^{2})^{3}+t_{5}\,x_{\mu}x_{\nu}x_{\lambda}x_{\sigma}/(x^{2})^{4}.
\end{align}
This is the most general tensor structure consistent with Lorentz invariance
and three permutation symmetries 1) $\mu\leftrightarrow\nu$; 2) $\lambda
\leftrightarrow\sigma$; 3) $\mu\nu\leftrightarrow\lambda\sigma$,
$x\rightarrow-x$.\footnote{This form is also valid in presence of parity
breaking, since in $D=2$ it turns out impossible to write down a parity
breaking term consistent with these symmetries.} Scaling is fixed from the
canonical dimension $[T_{\mu\nu}]=D$. We assume that there are no logarithms,
see below. It will be convenient to use an alternative parametrization in
terms of derivatives:
\begin{align}
\left\langle T_{\mu\nu}(x)T_{\lambda\sigma}(0)\right\rangle  &  =[a_{1}%
\,(\eta_{\mu\lambda}\eta_{\nu\sigma}+\eta_{\nu\lambda}\eta_{\mu\sigma}%
)+a_{2}\,\eta_{\mu\nu}\eta_{\lambda\sigma}]/(x^{2})^{2}\nonumber\\
&  +a_{3}\,(\eta_{\mu\lambda}\partial_{\nu}\partial_{\sigma}\frac{1}{x^{2}%
}+\text{3 perms})+a_{4}\,(\eta_{\mu\nu}\partial_{\lambda}\partial_{\sigma
}\frac{1}{x^{2}}+\eta_{\lambda\sigma}\partial_{\mu}\partial_{\nu}\frac
{1}{x^{2}})\nonumber\\
&  +a_{5}\,\partial_{\mu}\partial_{\lambda}\partial_{\sigma}\partial_{\nu}%
\log(x^{2})\,,\label{eq:TTx}%
\end{align}
where the constants $a_{i}$ are certain linear combinations of $t_{i}$.
Passing to Fourier transform we get:%
\begin{align}
\left\langle T_{\mu\nu}(k)T_{\lambda\sigma}(-k)\right\rangle  &  =\,\log
k^{2}\Bigl[A_{1}k^{2}(\eta_{\mu\lambda}\eta_{\nu\sigma}+\eta_{\nu\lambda}%
\eta_{\mu\sigma})+A_{2}k^{2}\eta_{\mu\nu}\eta_{\lambda\sigma}\label{eq:TT}\\
&  +A_{3}(\eta_{\mu\lambda}k_{\nu}k_{\sigma}+\text{3 perms)}+A_{4}(\eta
_{\mu\nu}k_{\lambda}k_{\sigma}+\eta_{\lambda\sigma}k_{\mu}k_{\nu}%
)\Bigr]+A_{5}k_{\mu}k_{\nu}k_{\lambda}k_{\sigma}/k^{2}\,\text{,}\nonumber
\end{align}
where $A_{i}\propto a_{i}$. Notice that even though $\log k^{2}$ is present,
the correlator changes only by a local quantity if the scale of the logarithm
is changed.

Now let us impose the conservation:%
\begin{align}
0=k^{\mu}\left\langle T_{\mu\nu}T_{\lambda\sigma}\right\rangle  &  =\,\log
k^{2}\Bigl[(A_{1}+A_{3})k^{2}(k_{\lambda}\eta_{\nu\sigma}+k_{\sigma}\eta
_{\nu\lambda})+(A_{2}+A_{4})k^{2}k_{\nu}\eta_{\lambda\sigma}\nonumber\\
&  +(2A_{3}+A_{4})k_{\nu}k_{\lambda}k_{\sigma}\Bigr]+A_{5}k_{\nu}k_{\lambda
}k_{\sigma}\,.
\end{align}
This equation does not constrain $A_{5}$ since its contribution to
$\partial^{\mu}\left\langle T_{\mu\nu}(x)T_{\lambda\sigma}(0)\right\rangle $
is purely local. On the other hand, the coefficient multiplying $\log k^{2}$
must be set to zero. The three terms making it up have different tensor
structure; their linear independence can be checked by going to the frame
$k^{\mu}=(1,0)$. We conclude that
\begin{equation}
A_{1}+A_{3}=A_{2}+A_{4}=2A_{3}+A_{4}=0\,, \label{eq:constr2D}%
\end{equation}
from where all the coefficients can be fixed in terms of, say, $A_{1}$, apart
from $A_{5}$ which is left undetermined.

This is all that can be derived on the basis of conservation alone. Now let us
see what this tells us about the 2-point function of the trace. From
(\ref{eq:TT}), we have
\begin{equation}
\left\langle T_{\mu}^{\mu}T_{\lambda}^{\lambda}\right\rangle =4(A_{1}%
+A_{2}+A_{3}+A_{4})\,k^{2}\log k^{2}+A_{5}k^{2}=0+\text{local}\,.
\end{equation}
Thus, the 2-point function of $T_{\mu}^{\mu}$ is zero up to local terms. In a
unitary theory (or reflection-positive, if we are working in the Euclidean),
this implies that $T_{\mu}^{\mu}\equiv0$, and thus the theory is conformal.

Let us list here two implicit assumptions of the given proof: 1) the stress
tensor 2-point function exists; 2) the stress tensor has canonical scaling,
without logarithms.

Assumption 2) can be expressed formally as the requirement that the commutator
with the scale generator $S$ take its canonical form:%
\begin{equation}
i\left[  S,T_{\mu\nu}(x)\right]  =x^{\rho}\partial_{\rho}T_{\mu\nu
}(x)+D\,T_{\mu\nu}(x)\,. \label{canonicalScal}%
\end{equation}
In general, one can add to the right-hand side a term $\partial^{\sigma
}\partial^{\rho}\tilde{Y}_{\mu\sigma\nu\rho}$, where $\tilde{Y}_{\mu\sigma
\nu\rho}$ has the same symmetry as $Y_{\mu\sigma\nu\rho}$ in (\ref{eq:EMmodif}%
). This would be consistent with the integrated relation $i[S,H]=H$, where $H$
is the Hamiltonian. However, the correlators of such $T_{\mu\nu}$ would in
general contain logarithms.

In this case one looks for a redefined, equivalent stress tensor with
canonical scaling. In \cite{Polchinski:1987dy}, it was shown that such a
redefinition can always be achieved provided that the theory has a discrete
spectrum of scaling dimensions. The redefined stress tensor 2-point function
is free of logarithms, and the above argument is applicable. A nice example of
this phenomenon can be found in Ref.~\cite{Riva:2005gd}. One of several
equivalent stress tensors considered in that paper, Eq.~(14), does not scale
canonically, and its trace 2-point function is nonzero. However, an
appropriate improvement exists which restores the canonical scaling, and leads
to the vanishing 2-point function of the trace. [The last step of the proof,
concluding that $T_{\mu}^{\mu}\equiv0$, cannot be carried out since the theory
of Ref.~\cite{Riva:2005gd} is not reflection-positive.]

Let us come back to the implicit assumption 1). Hull and Townsend
\cite{Hull:1985rc} have shown that scale-invariant but not
conformally-invariant unitary theories exist among 2D sigma-models with
non-compact target space. These models likely violate assumption 1), and
perhaps also 2) \cite{Polchinski:1987dy}.

\subsection{$D\geq3$: extra ideas needed}

\label{sec:extra}

In Appendix A, we repeat the $D=2$ argument in higher dimensions and see that
it does not go through. In other words, in $D\geq3$ it is impossible to
conclude from scaling, conservation, and unitarity alone that the stress
tensor is traceless. In fact, we can understand this by an explicit example.
Consider the free massless scalar theory. Its non-improved stress tensor,%
\[
T_{\mu\nu}=\partial_{\mu}\phi\partial_{\nu}\phi-\frac{1}{2}\eta_{\mu\nu
}(\partial\phi)^{2}\,,
\]
is conserved, the 2-point function is unitary, and does not contain
logarithms. Yet $T_{\mu}^{\mu}=(1-D/2)(\partial\phi)^{2}$ is not vanishing for
$D\geq3$. Of course we know that in this case an improved traceless tensor
exists. But if you are given a random stress tensor you cannot expect to be
able to prove that it's traceless.

What this means is that in $D\geq3$ we should start from (\ref{eq:TK}) and aim
for proving (\ref{eq:KL}). This is weaker than tracelessness, but is enough to
show that the theory is conformal.

Why is there such a difference between $D\geq3$ and $D=2$? One explanation is
as follows. In $D=2$, among all the equivalent stress tensors, only one will
have the 2-point function which scales canonically, without logarithms. This
is because $Y_{\mu\rho\nu\sigma}$ is dimensionless in 2D, and its 2-point
function is logarithmic. Then, it turns out that this very special
canonically-scaling $T_{\mu\nu}$ is traceless. In $D\geq3$ the situation is
different, since $Y_{\mu\rho\nu\sigma}$ has positive mass dimension. Thus the
transformation (\ref{eq:EMmodif}) does not have to introduce logarithms.

At present, one can only hypothesize on how (\ref{eq:KL}) could be derived.
For example, one could start with correlators containing one $T_{\mu\nu}$
insertion and some other fields of the theory. These correlators are
constrained via Ward identities. Another constraint is provided by
(\ref{eq:TK}). Now, perhaps one could show that correlators of $K_{\mu}$
satisfy an integrability condition which allows to define a local field
$L_{\mu\nu}$, consistently with (\ref{eq:KL}). Then one would have to extend
the definition of $L_{\mu\nu}$ to correlators with two stress tensor
insertions etc. This is not an easy plan to carry out. Also, we know that
unitarity must enter the scene, and it is not clear how this will happen.

\section{Looking for a counterexample}

\label{sec:count}

It makes sense to attack the problem from both sides. If the theorem is not
true, a counterexample must exist. In this section we will discuss what is
known in this direction. We will explain that one should restrict attention to
theories which have stress tensor. Actually, if this physical requirement is
dropped, the problem trivializes and simple counterexamples exist. We will
also rule out the presence of counterexamples within a new large class of
models in $4-\varepsilon$ dimensions.

\subsection{Fake counterexamples without stress tensor}

First a general remark about theories which \textit{should not} be considered
as counterexamples. We are referring to field theories which, while being
unitary and scale invariant, do not have a stress tensor. As is well known,
Wightman axioms require existence of the full energy and momentum, but not of
their density, which is the stress tensor. However, it seems reasonable to
adopt existence of a stress tensor as an additional axiom that a full-fledged
field theory must satisfy. Otherwise we would not know how to couple the
theory to (classical) gravity.

Consider a theory which contains a vector field $A_{\mu}$ of scaling dimension
$\Delta$, whose 2-point function is given by:%
\begin{equation}
\left\langle A_{\mu}(x)A_{\nu}(0)\right\rangle =(x^{2})^{-\Delta}(\alpha
\eta_{\mu\nu}-2x_{\mu}x_{\nu}/x^{2})\,.\label{eq:AA}%
\end{equation}
We assume that the theory is free (Gaussian). In practice this means that all
higher-order correlators are computed from (\ref{eq:AA}) via Wick's theorem.
Composite fields can be defined via OPE. In essence, this defines a field
theory, which is unitary if and only if the two-point function (\ref{eq:AA})
is unitary.

One can analyze the unitarity of (\ref{eq:AA})\ by the same method we used to
study the stress tensor 2-point function in Appendix A, see also
\cite{Grinstein:2008qk}. The answer is that it is unitary if and only if%
\begin{equation}
\Delta\geq2\text{,\quad}1/\Delta\leq\alpha\leq2-3/\Delta\quad(D=4)\,.
\label{eq:Aun}%
\end{equation}

Now, let us analyze when the theory based on (\ref{eq:AA}) is conformal. This
can happen in only two ways: 1) $A_{\mu}$ is a descendant of a primary scalar
field, $A_{\mu}=\partial_{\mu}\phi$; 2) $A_{\mu}$ is a primary vector field.
Case 1) is realized for $\alpha=1/\Delta$, at the lower bound of the interval
allowed by untarity. Case 2) requires $\alpha=1$, which is consistent with
unitarity provided $\Delta\geq3$, a well-known result \cite{mack}.

Any other value of $\alpha$ will give a theory which is scale invariant,
unitary, but not conformal.

As mentioned above, we consider this a fake counterexample because this theory
does not have a stress tensor. In the conformal case, one could introduce a
stress tensor in the $1/N$ expansion realizing this model via AdS/CFT. In the
non-conformal case, we do not know how to do even that, without introducing
pathologies or lowering cutoff in the dual gravitational theory.

Another fake counterexample is linearized gravity\footnote{See \cite{anselmi}
for a discussion; we are grateful to Damiano Anselmi for bringing this example
to our attention.}. Again, there is no (gauge-invariant) stress tensor in this
theory. This can be viewed as a consequence of the Weinberg-Witten theorem
\cite{Weinberg:1980kq}.

\subsection{Systematic search in $4-\varepsilon$ dimensions}

As explained in the introduction, a putative counterexample theory must, to
begin with, contain a non-conserved, dimension 3, hermitean vector field
$K_{\mu}$ which can appear in (\ref{eq:TK}). Nontrivial $K_{\mu}$ candidates
should not be expressible as in (\ref{eq:KL}), since in the latter case the
theory is conformal,

The simplest model with $K_{\mu}$ candidates was considered in
\cite{Polchinski:1987dy}; it is a theory of $N$ massless real scalars with
quartic self-interaction, Eq.~(\ref{eq:phi4}). The nontrivial $K^{\mu}$
candidates are given by (\ref{eq:Kmu}). The $Q_{[ij]}$ is a real matrix,
assumed antisymmetric since otherwise $K_{\mu}$ is a total derivative, which
is a partial case of (\ref{eq:KL}).

\textit{\`{A} la} Wilson-Fisher, the model can have fixed points in
$4-\varepsilon$ dimensions. The one-loop $\beta$-function is given by%
\begin{equation}
\beta_{ijkl}=-\epsilon\lambda_{ijkl}+\frac{1}{16\pi^{2}}(\lambda_{ijmn}%
\lambda_{mnkl}+2\text{ perms})\,.
\end{equation}
We assume that $\lambda_{ijkl}$ is symmetrized; it is also real as required by
unitarity. The condition%
\begin{equation}
\beta=0 \label{eq:conf}%
\end{equation}
gives a scale invariant theory which is also conformally invariant, since at
one-loop the trace anomaly is given by%
\begin{equation}
T_{\mu}^{\mu}=\frac{1}{4!}\beta_{ijkl}s_{i}s_{j}s_{k}s_{l}\,.
\end{equation}
However, Eq.~(\ref{eq:conf}) is not the most general condition for scale
invariance. The $\beta$-function encodes a change in the couplings when we
integrate out a momentum shell. The theory will remain scale invariant if this
change, though nonzero, can be compensated by adding to the Lagrangian a total
derivative term $\partial_{\mu}K^{\mu}$. Using equations of motion (EOM) at
leading order in the coupling, this term can be rewritten as%
\begin{equation}
\partial_{\mu}K^{\mu}\cong\frac{1}{4!}\mathcal{Q}_{ijkl}s_{i}s_{j}s_{k}%
s_{l},\qquad\mathcal{Q}_{ijkl}=Q_{[im]}\lambda_{mjkl}+3\text{ perms.}\,
\end{equation}
Thus the most general condition for a scale invariant fixed point:%
\begin{equation}
T_{\mu}^{\mu}=\partial_{\mu}K^{\mu}\quad\Longleftrightarrow\quad
\beta=\mathcal{Q}. \label{eq:sc}%
\end{equation}

A fixed point (\ref{eq:sc}) would break conformal invariance if $\mathcal{Q}%
\neq0$. However, as noticed in \cite{Polchinski:1987dy}, this never happens.
To see this, we contract with $\mathcal{Q}$ and get:%
\begin{equation}
\mathcal{Q}\cdot\mathcal{Q=\,}\beta\cdot\mathcal{Q}\equiv0\,, \label{eq:QQ}%
\end{equation}
where the RHS vanishes identically by the symmetry properties of
$\lambda_{ijkl}$ and $Q_{[ij]}$:%
\begin{align}
\lambda_{Ijkl}Q_{[IM]}\lambda_{Mjkl}  &  =0\,,\nonumber\\
(\lambda_{Ijmn}\lambda_{mnkl}+\lambda_{Ikmn}\lambda_{mnjl}+\lambda
_{Ilmn}\lambda_{mnjk})Q_{[IM]}\lambda_{Mjkl}  &  =3Q_{[IM]}\lambda
_{Ijmn}\lambda_{Mjkl}\lambda_{mnkl}=0\,. \label{eq:manip}%
\end{align}
[In both cases, antisymmetric $Q_{[IM]}$ is contracted with a symmetric
tensor.]\ Eq.~(\ref{eq:QQ}) implies that $\mathcal{Q}\equiv0$ and we are back
to the conformally-invariant case (\ref{eq:conf}).

It is not immediately clear what to make out of this result. On the one hand,
the considered class of models was rather large. On the other hand, it is by
far not the most general unitary field theory which can have perturbative
fixed points in $D=4-\varepsilon$. It seems reasonable to try to complicate
the model, in the hope that a new qualitative effect may appear. For example,
reality of couplings seems to have played a role in the above argument, and
one could think that breaking CP may help in constructing a counterexample.

For this reason, we extend the field content of the model, by adding an
arbitrary number of Weyl fermions, and making them interact with the scalars
via Yukawa couplings. The full Lagrangian is now the sum of (\ref{eq:phi4})
and (\ref{eq:psi}). In general, it breaks both CP and P. If these discrete
symmetries have anything to do with the relation between scale and conformal
invariance, we can hope to detect this.

Let us now repeat the steps of the above analysis for the new theory. The
standard $\beta$-functions are given by:%
\begin{align}
\beta_{ijkl}^{(\lambda)}= &  -\epsilon\lambda_{ijkl}+\frac{1}{16\pi^{2}%
}[\lambda_{ijmn}\lambda_{mnkl}+2\text{ perm}]+\frac{1}{4\pi^{2}}%
[\text{Tr}(y_{i}^{\ast}y_{m}+y_{i}y_{m}^{\ast})\lambda_{mjkl}+\text{3
perms}]\nonumber\\
&  -\frac{1}{4\pi^{2}}[\text{Tr}(y_{i}y_{j}^{\ast}y_{k}y_{l}^{\ast}%
+y_{i}^{\ast}y_{j}y_{k}^{\ast}y_{l})+5\text{ other perms }%
(jkl)]\,,\label{eq:bb}\\
\beta_{i|ab}^{(y)}= &  -\frac{\epsilon}{2}y_{i|ab}+\frac{1}{4\pi^{2}}%
\text{Tr}(y_{i}^{\ast}y_{j}+y_{i}y_{j}^{\ast})y_{j|ab}+\frac{1}{8\pi^{2}%
}(y_{j}y_{i}^{\ast}y_{j})_{ab}\,+\frac{1}{8\pi^{2}}\left[  (y_{i}(y_{j}%
y_{j}^{\ast}))_{ab}+((y_{j}y_{j}^{\ast})y_{i})_{ab}\right]  .\nonumber
\end{align}
The matrix $y_{i|ab}\equiv(y_{i})_{ab}$ is assumed symmetrized in $ab$. Matrix
multiplication and trace are in the fermion flavor space.

Now, the usual fixed points satisfying
\begin{equation}
\beta^{(\lambda)}=0,\quad\beta^{(y)}=0\text{,}%
\end{equation}
will have both scale and conformal invariance. More general fixed points are
associated with nontrivial $K_{\mu}$ candidates, which in this model have the
form
\begin{equation}
K_{\mu}=Q_{[ij]}\,s_{i}\partial_{\mu}s_{j}\,+iP_{ab}\,{\bar{\psi}}_{a}%
{\bar{\sigma}}_{\mu}\psi_{b}\,.\label{FermCurr1}%
\end{equation}
Here $P$ must be antihermitean to have a hermitean $K_{\mu}$.

As before, we try to compensate the renormalization group transformation by
adding the total derivative $\partial_{\mu}K^{\mu}$ and re-expressing via the
EOM. We have%
\begin{equation}
\partial_{\mu}K^{\mu}\cong\frac{1}{4!}\mathcal{Q}_{ijkl}s_{i}s_{j}s_{k}%
s_{l}+\frac{1}{2!}(\mathcal{P}_{i|ab}s_{i}\psi_{a}\psi_{b}+\text{h.c.}),
\end{equation}
where $\mathcal{Q}$ is the same as before, and
\begin{equation}
\mathcal{P}_{i|ab}=Q_{ij}y_{j|ab}+[(y_{i}P)_{ab}+a\leftrightarrow b]\,.
\end{equation}

Thus, fixed points without conformal invariance are the solutions of
\begin{equation}
\beta^{(\lambda)}=\mathcal{Q},\quad\beta^{(y)}=\mathcal{P}\text{\thinspace}
\label{eq:bQbP}%
\end{equation}
with nonzero $\mathcal{P}$ and/or $\mathcal{Q}$. As we will show now, no such
solutions exist.

\textbf{Theorem} \textit{All solutions of this equation have zero
}$\mathcal{P}$\textit{ and }$\mathcal{Q}$\textit{, and thus do not break
conformal symmetry.}

The proof is based on the same idea. We contract (\ref{eq:bQbP}) with
$\mathcal{Q}$ and $\mathcal{P}^{\ast}$ and show that $\beta^{(\lambda)}%
\cdot\mathcal{Q}$ and $\beta^{(y)}\cdot\mathcal{P}^{\ast}$ vanish. This way we
conclude first that $\mathcal{P}=0$, and then that $\mathcal{Q}=0$. It turns
out crucial to proceed in this order because, as we will see, $\beta
^{(\lambda)}\cdot\mathcal{Q}$ does not vanish identically but only modulo
terms proportional to $\mathcal{P}$.

These statements can be and were verified by tensor manipulations analogous to
(\ref{eq:manip}), only more tedious. We will now show an alternative,
diagrammatic, way of organizing this computation. The $\beta$-functions
(\ref{eq:bb}) are the sum of the following Feynman graphs:%
\begin{equation}
\beta^{(\lambda)}=%
\raisebox{-0.2572in}{\includegraphics[
height=0.6262in,
width=0.7122in
]%
{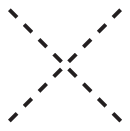}%
}%
+%
\raisebox{-0.2572in}{\includegraphics[
height=0.5872in,
width=1.3172in
]%
{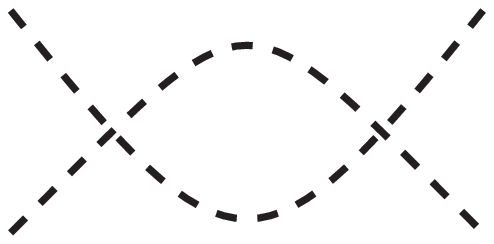}%
}%
+%
\raisebox{-0.2572in}{\includegraphics[
height=0.6821in,
width=0.7335in
]%
{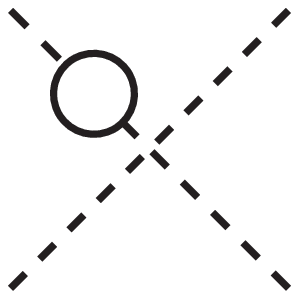}%
}%
+%
\raisebox{-0.2572in}{\includegraphics[
height=0.6821in,
width=0.7424in
]%
{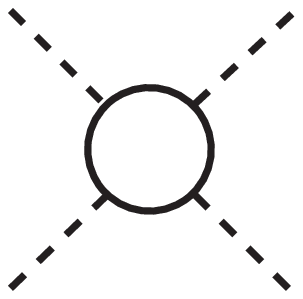}%
}%
+\text{perms\thinspace,} \label{eq:gbl}%
\end{equation}%
\begin{equation}
\beta^{(y)}=%
\raisebox{-0.2572in}{\includegraphics[
height=0.5748in,
width=0.6759in
]%
{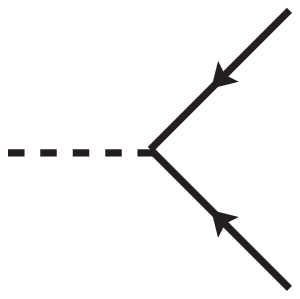}%
}%
+%
\raisebox{-0.2572in}{\includegraphics[
height=0.5748in,
width=0.8036in
]%
{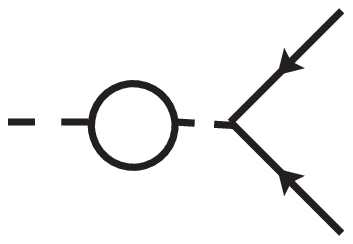}%
}%
+%
\raisebox{-0.3087in}{\includegraphics[
height=0.7468in,
width=0.8036in
]%
{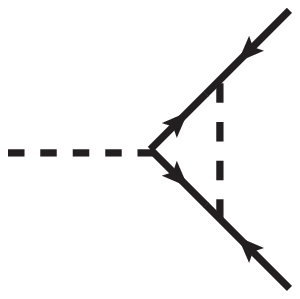}%
}%
+%
\raisebox{-0.3087in}{\includegraphics[
height=0.8089in,
width=0.9305in
]%
{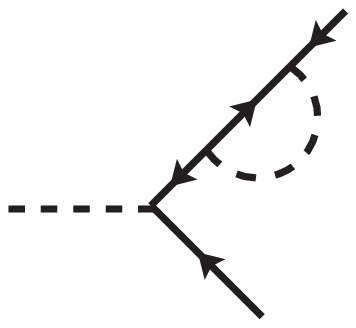}%
}%
\text{+ perm\thinspace.}%
\end{equation}
with the vertices%
\begin{equation}
\lambda_{ijkl}=%
\raisebox{-0.2572in}{\includegraphics[
height=0.6262in,
width=0.7122in
]%
{l.eps}%
}%
,\quad y_{i|ab}=%
\raisebox{-0.2572in}{\includegraphics[
height=0.5748in,
width=0.6759in
]%
{f.eps}%
}%
,\quad y_{i|ab}^{\ast}=%
\raisebox{-0.2572in}{\includegraphics[
height=0.5748in,
width=0.6759in
]%
{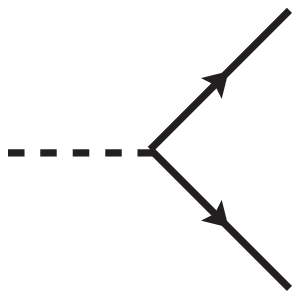}%
}%
\,.
\end{equation}
The arrows on fermion lines show the flow of chirality. If no arrows are shown
in a fermion loop, a sum over both ways to distribute the arrows is presumed.

The precise numerical values of the loop integrals, appearing as prefactors in
(\ref{eq:bb}), will not be important below. Thus we strip Feynman graphs from
all spacetime dependence (propagators, $\sigma$-matrices). The only thing that
counts is that these graphs correctly encode the tensor structure of various
terms in (\ref{eq:bb}).

The $\mathcal{Q}$ and $\mathcal{P}$ tensors can be encoded in the same
language, introducing new vertices to denote contractions with $Q$ and $P$:%
\begin{equation}
\mathcal{Q}=%
\raisebox{-0.2572in}{\includegraphics[
height=0.6537in,
width=0.8914in
]%
{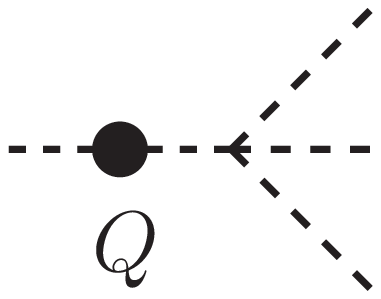}%
}%
+\text{perms\thinspace,} \label{eq:gQ}%
\end{equation}%
\begin{equation}
\mathcal{P}=%
\raisebox{-0.3087in}{\includegraphics[
height=0.6821in,
width=0.9056in
]%
{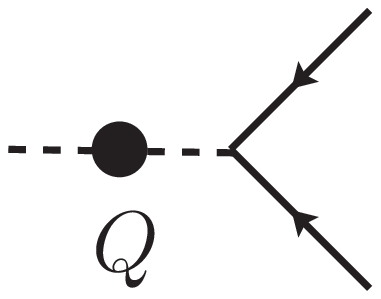}%
}%
+%
\raisebox{-0.3087in}{\includegraphics[
height=0.738in,
width=0.9908in
]%
{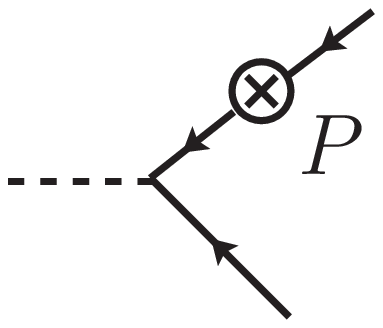}%
}%
+\text{perm\thinspace.}%
\end{equation}

Let us use this formalism to give a diagrammatic proof of (\ref{eq:QQ}). We
have to contract the $\mathcal{Q}$ graph (\ref{eq:gQ}) with the purely scalar
diagrams in the graphical representation of $\beta^{(\lambda)}$,
Eq.~(\ref{eq:gbl}). This contraction gives diagrams of two types:%

\begin{equation}%
\raisebox{-0.2572in}{\includegraphics[
height=0.84in,
width=1.0484in
]%
{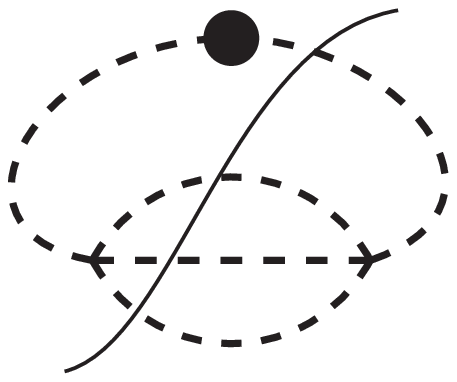}%
}%
\qquad%
\raisebox{-0.2271in}{\includegraphics[
height=0.7797in,
width=1.0821in
]%
{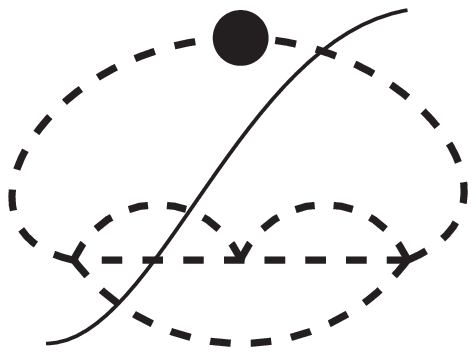}%
}%
\,. \label{eq:Q12}%
\end{equation}
The thin wavy lines cutting the diagrams show where indices were contracted.

Both these diagrams look like a contraction of (antisymmetric) $Q$ with a
left-right symmetric graph. Clearly, these contractions are zero. Two graphs
(\ref{eq:Q12}) neatly visualize the content of Eqs. (\ref{eq:manip}).

Armed with this formalism, we will now prove the theorem without danger of
getting lost in the forest of indices. We begin by contracting the second
Eq.~(\ref{eq:bQbP}) with $\mathcal{P}^{\ast}$:%
\begin{equation}
\mathcal{P}\cdot\mathcal{P}^{\ast}=\beta^{(y)}\cdot\mathcal{P}^{\ast}\,.
\label{eq:PP}%
\end{equation}
Various terms appearing in the RHS correspond to contractions of $Q$ and $P$
and are as follows:
\begin{equation}%
\raisebox{-0.2572in}{\includegraphics[
height=0.7309in,
width=0.8488in
]%
{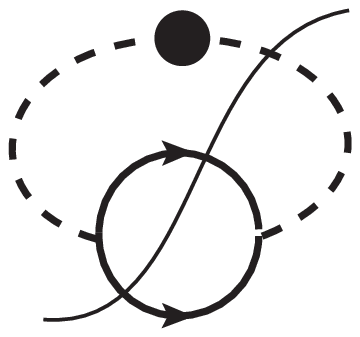}%
}%
\qquad%
\raisebox{-0.2572in}{\includegraphics[
height=0.6741in,
width=1.0404in
]%
{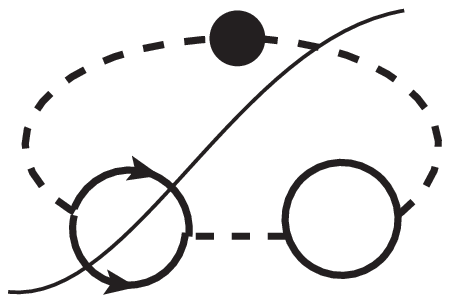}%
}%
\qquad%
\raisebox{-0.2572in}{\includegraphics[
height=0.7539in,
width=0.926in
]%
{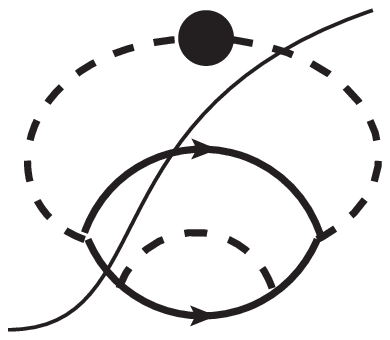}%
}%
\qquad%
\raisebox{-0.2572in}{\includegraphics[
height=0.7539in,
width=0.926in
]%
{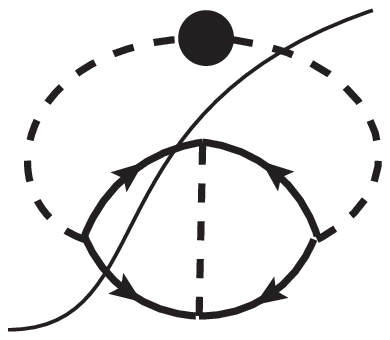}%
}%
\end{equation}%
\begin{equation}%
{\includegraphics[
height=0.7601in,
width=0.8763in
]%
{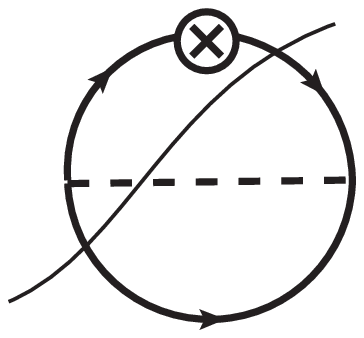}%
}%
\qquad%
{\includegraphics[
height=0.7965in,
width=0.8116in
]%
{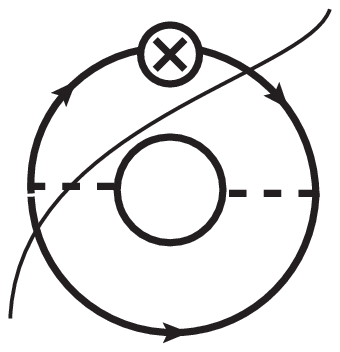}%
}%
\qquad%
{\includegraphics[
height=0.7965in,
width=0.8036in
]%
{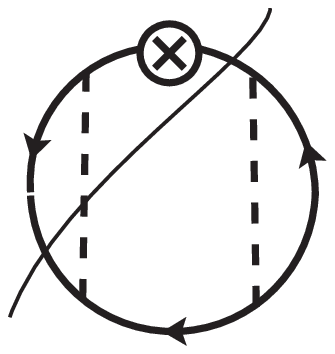}%
}%
\qquad%
{\includegraphics[
height=0.7965in,
width=0.8036in
]%
{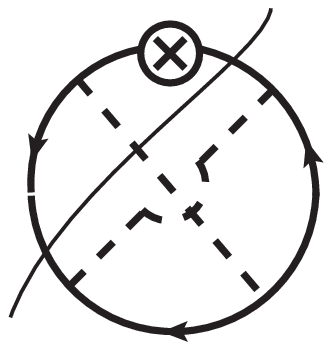}%
}%
\,.
\end{equation}
All these graphs are symmetric, up to inversion of the arrow directions, which
corresponds to complex conjugation. This means that the real antisymmetric $Q$
and antihermitean $P$ are contracted with hermitean matrices. Such
contractions are pure imaginary, and do not contribute to the manifestly real
LHS of (\ref{eq:PP}).

We conclude that $\mathcal{P}\cdot\mathcal{P}^{\ast}=0$, and thus
$\mathcal{P}=0$. Let us proceed with $\mathcal{Q}$. From the first
Eq.~(\ref{eq:bQbP}) we have:%
\begin{equation}
\mathcal{Q}\cdot\mathcal{Q}=\beta^{(\lambda)}\cdot\mathcal{Q}\,.
\end{equation}
Purely scalar terms in the RHS were already shown above to vanish identically.
The terms involving fermion loops are:%
\begin{equation}%
{\includegraphics[
height=0.8763in,
width=1.0493in
]%
{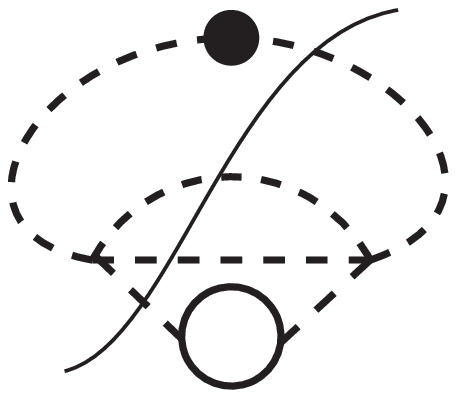}%
}%
\qquad%
{\includegraphics[
height=0.8373in,
width=1.0821in
]%
{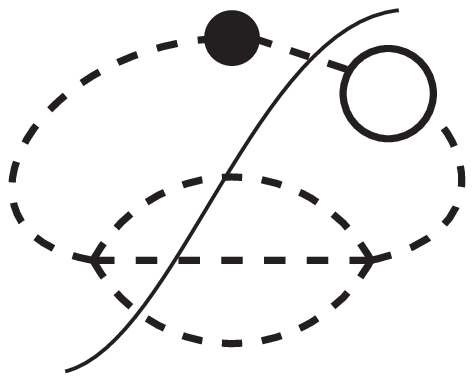}%
}%
\qquad%
{\includegraphics[
height=0.8373in,
width=1.0661in
]%
{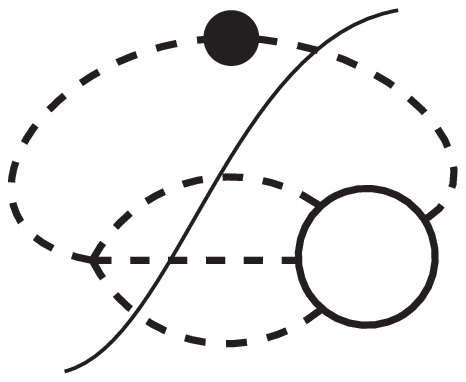}%
}%
\,.
\end{equation}
The first graph is symmetric and vanishes for the usual reason. The other two
do not exhibit any obvious symmetry. However, let us invoke the already shown
property $\mathcal{P}=0$. Graphically:%
\begin{equation}%
\raisebox{-0.2572in}{\includegraphics[
height=0.6821in,
width=0.9056in
]%
{P1.eps}%
}%
=-\left(
\raisebox{-0.3087in}{\includegraphics[
height=0.738in,
width=0.9908in
]%
{P2.eps}%
}%
+\text{perm}\right)  \,.
\end{equation}
Using this identity, the two non-symmetric graphs are transformed into:%
\begin{equation}%
{\includegraphics[
height=0.8967in,
width=1.0049in
]%
{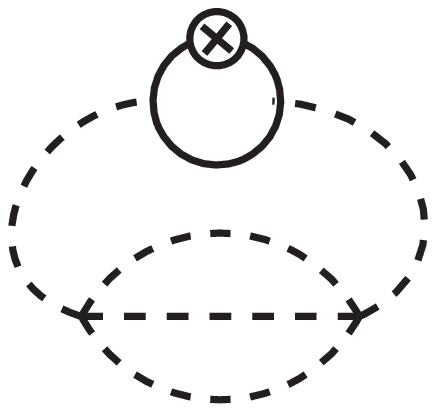}%
}%
\qquad%
{\includegraphics[
height=0.7344in,
width=0.5898in
]%
{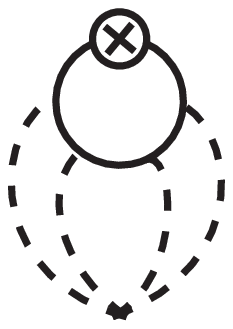}%
}%
,
\end{equation}
which \textit{are} symmetric and vanish. The proof is complete.

\section{Conclusions}

\label{concl}

In this paper we gave a fresh look at the relation of scale and conformal
invariance in unitary field theories in $D>2.$ As we discussed at length, the
situation is frustrating: we don't know if one implies the other, although
this holds in all known examples. We added to the list a new large class of
models: multi-field theories of massless scalars and Weyl fermions with
arbitrary quartic and Yukawa interactions. What makes these models interesting
is that they contain nontrivial candidates for a non-conserved dimension 3
vector which could in principle appear as a correction term in the scale
current, thus making a scale-invariant theory non-conformal. However, we
showed that this never happens for the one-loop fixed points in $4-\varepsilon
$ dimensions. Our current proof proceeds by inspecting symmetry properties of
certain Feynman graphs. We suspect that a deeper explanation must exist and
hope that it may one day be found.

In conclusion, we would like to point out that if scale-invariant but
non-conformal unitary theories exist, this may have not only theoretical but
also phenomenological consequences:

\begin{enumerate}
\item In the unparticle physics scenario \cite{Georgi} one assumes the
existence of a scale-invariant hidden sector weakly coupled to the Standard
Model via non-renormalizable operators. To study phenomenology of such a
scenario, one would like to know if dimensions and propagators of vector
operators in the hidden sector must follow the rules of conformal theories, or
can be more general \cite{Grinstein:2008qk}.

\item In the conformal technicolor scenario of the electroweak symmetry
breaking \cite{luty}, one assumes that the Higgs field $H$ belongs to a
strongly interacting sector with conformal invariance above a TeV.
Furthermore, one assumes an unusual pattern of operator dimensions: $H$ should
have dimension close to $1$ (to avoid problems with flavor), while the
composite operator $H^{\dagger}H$ must have dimension close to $4$ (to solve
the hierarchy problem). It is not known if such large deviation from the naive
relation $[H^{\dagger}H]\simeq2[H]$ can be realized in a conformal field
theory. In fact, recent work \cite{r} uses conformal symmetry to show that
there is least one scalar in the OPE $H^{\dagger}\times H,$ whose dimension is
close to $2$ if $[H]\simeq1$. While it is not yet known if this scalar is a
singlet $H^{\dagger}H$ or a triplet $H^{\dagger}\sigma^{a}H$, one can well
imagine that future studies may rule out conformal technicolor. In this case
it would be interesting to know if the scenario could be saved by assuming
that the Higgs sector is scale-invariant but not conformal, in which case the
bounds of \cite{r}\ do not apply.
\end{enumerate}

\section*{Acknowledgements}

We are grateful to Jan Troost for useful discussions and comments. This work
was partially supported by the EU under RTN contract MRTN-CT-2004-503369 and
by MIUR under the contract PRIN-2006022501.

\appendix

\section{Stress tensor two-point function in $D\geq3$}

In this section, we will explore how conservation and unitarity constrain the
stress tensor 2-point function in $D\geq3$. As explained in Section
\ref{sec:extra}, we cannot expect to derive tracelessness. It is instructive
to see where exactly the $D=2$ argument breaks down.

Analogously to (\ref{eq:TTx}), we have%
\begin{align}
\left\langle T_{\mu\nu}(x)T_{\lambda\sigma}(0)\right\rangle  &  =[a_{1}%
\,(\eta_{\mu\lambda}\eta_{\nu\sigma}+\eta_{\nu\lambda}\eta_{\mu\sigma}%
)+a_{2}\,\eta_{\mu\nu}\eta_{\lambda\sigma}]/(x^{2})^{D}\nonumber\\
&  +a_{3}\,(\eta_{\mu\lambda}\partial_{\nu}\partial_{\sigma}\frac{1}%
{(x^{2})^{D-1}}+\text{3 perms})+a_{4}\,(\eta_{\mu\nu}\partial_{\lambda
}\partial_{\sigma}\frac{1}{(x^{2})^{D-1}}+\eta_{\lambda\sigma}\partial_{\mu
}\partial_{\nu}\frac{1}{(x^{2})^{D-1}})\nonumber\\
&  +a_{5}\,\partial_{\mu}\partial_{\lambda}\partial_{\sigma}\partial_{\nu
}\frac{1}{(x^{2})^{D-2}}\,.\label{eq:4Dx}%
\end{align}
This is the most general form in $D=4$. In $D=3$ one could also consider a
parity-breaking term $\propto\varepsilon_{\mu\lambda\rho}x^{\rho}\eta
_{\nu\sigma}$+symmetrizations; we do not analyze this possibility.

Passing to the momentum space,%
\begin{align}
\left\langle T_{\mu\nu}(k)T_{\lambda\sigma}(-k)\right\rangle  &  =f_{D}%
(k^{2})\Bigl[A_{1}k^{2}(\eta_{\mu\lambda}\eta_{\nu\sigma}+\eta_{\nu\lambda
}\eta_{\mu\sigma})+A_{2}k^{2}\eta_{\mu\nu}\eta_{\lambda\sigma}\label{eq:4Dk}\\
&  +A_{3}(\eta_{\mu\lambda}k_{\nu}k_{\sigma}+\text{3 perms)}+A_{4}(\eta
_{\mu\nu}k_{\lambda}k_{\sigma}+\eta_{\lambda\sigma}k_{\mu}k_{\nu}%
)+A_{5}\,k_{\mu}k_{\nu}k_{\lambda}k_{\sigma}/k^{2}\Bigr]\,\text{,}\nonumber
\end{align}
where%
\begin{equation}
f_{3}\propto(k^{2})^{3/2},\quad f_{4}\propto(k^{2})^{2}\log k^{2}\,.\,
\end{equation}
We now see a crucial difference with $D=2$. In 2D, the $A_{5}$ 4-derivative
term in (\ref{eq:TTx}) had an analytic structure in the momentum space,
Eq.~(\ref{eq:TT}), different from the other terms. That's why it dropped out
from the conservation constraint and contributed only a local term to the
2-point function of the trace. In retrospect, this was essentially due to an
accident, that in 2D the number of indices of $T_{\mu\nu}$ becomes equal to
its scaling dimension, which led to the appearance of the logarithm in
Eq.~(\ref{eq:TTx}). This does not happen in $D\geq3$, where all 5 terms in
(\ref{eq:4Dx}) are on equal footing. Thus, we can already foresee that
conservation will not be as constraining.

To see this explicitly, we have:%
\begin{align}
0=k^{\mu}\left\langle T_{\mu\nu}T_{\lambda\sigma}\right\rangle  &
=f_{D}(k^{2})[(A_{1}+A_{3})(k_{\lambda}\eta_{\nu\sigma}+k_{\sigma}\eta
_{\nu\lambda})+(A_{2}+A_{4})k_{\nu}g_{\lambda\sigma}\nonumber\\
&  +(2A_{3}+A_{4}+A_{5})k_{\nu}k_{\lambda}k_{\sigma}/(k^{2})]\,,
\label{Trace4}%
\end{align}
and we conclude
\begin{equation}
A_{1}+A_{3}=A_{2}+A_{4}=2A_{3}+A_{4}+A_{5}=0\,, \label{eq:Constraint}%
\end{equation}
which is weaker than (\ref{eq:constr2D}).

Let us now see which further constraints on $A_{i}$ come out by imposing
unitarity. The unitarity constraint is obtained by considering the non
time-ordered 2-point function in Minkowski space. The spectral density is
given by the imaginary part of the Schwinger function in Fourier space
(\ref{eq:4Dk}), and is concentrated in the forward lightcone $k^{0}>0$,
$k^{2}\equiv-k_{0}^{2}+\vec{k}^{2}<0$. In a unitary theory, the tensorial
spectral density must be positive definite, i.e.~for any fixed tensor
$Y_{\mu\nu}$ the operator $Y^{\mu\nu}T_{\mu\nu}$ must have positive spectral density.

An important partial case is $Y_{\mu\nu}=\eta_{\mu\nu}$, which corresponds to
studying the 2-point function of the trace. From (\ref{eq:4Dk}),
(\ref{eq:Constraint}), we have%
\begin{equation}
\left\langle T_{\mu}^{\mu}(k)T_{\lambda}^{\lambda}(-k)\right\rangle
=(D-1)[2A_{1}+(D-1)A_{2}]\,f_{D}(k^{2})\,, \label{eq:2ptTr}%
\end{equation}
where we expressed all $A_{i}$ in terms of $A_{1},A_{2}$. Thus
\begin{equation}
T_{\mu}^{\mu}\text{ unitarity\quad}\Longleftrightarrow\quad2A_{1}%
+(D-1)A_{2}\geq0\,. \label{eq:Trun}%
\end{equation}

Passing to general $Y_{\mu\nu}$, by Lorentz invariance it is enough to examine
the spectral density for $\vec{k}=0$. Multiplying (\ref{eq:4Dk}) with a
symmetric $Y^{\mu\nu}$ and its conjugate, and extracting the imaginary part,
we get the following constraint on the coefficients:%
\begin{equation}
2A_{1}Y_{\mu\nu}^{\ast}Y^{\mu\nu}+A_{2}|Y_{\mu}^{\mu}|^{2}-4A_{3}Y_{\mu
0}^{\ast}Y_{0}^{\mu}-A_{4}(Y_{\mu}^{\mu}Y_{00}^{\ast}+\text{h.c.}%
)+A_{5}\,|Y_{00}|^{2}\geq0\,.
\end{equation}
Separating the spatial and temporal coordinates, this condition can be
rewritten as:%
\begin{align}
&  (2A_{1}+A_{2}+4A_{3}+2A_{4}+A_{5})|Y_{00}|^{2}-4(A_{1}+A_{3})|Y_{0i}%
|^{2}-(A_{2}+A_{4})(Y_{ii}Y_{00}^{\ast}+\text{h.c.})\nonumber\\
&  +2A_{1}Y_{ij}^{\ast}Y_{ij}+A_{2}|Y_{ii}|^{2}\,\geq0,
\end{align}
The first three terms drop out thanks to the conservation constraints
(\ref{eq:Constraint}), and we are left with%
\begin{equation}
2A_{1}Y_{ij}^{\ast}Y_{ij}+A_{2}|Y_{ii}|^{2}\,\geq0\,. \label{eq:Yij}%
\end{equation}
The necessary and sufficient conditions for this to be true are derived by
considering separately off-diagonal and diagonal $Y_{ij}$. We obtain:%
\begin{equation}
T_{\mu\nu}\text{ unitarity\quad}\Longleftrightarrow\quad A_{1}\geq
0\,,\quad2A_{1}+(D-1)A_{2}\geq0\,.
\end{equation}
This is not much extra mileage compared to the partial case (\ref{eq:Trun}%
)\footnote{Cardy \cite{Cardy:1988cwa} considers reflection positivity for the
$T_{\mu\nu}$ components in $D\geq3$ and states that the trace 2-point function
gives the strongest condition. However, we do find an extra condition
$A_{1}\geq0$.}. As expected, we cannot conclude that $T_{\mu}^{\mu}=0$.

\end{document}